# A Novel Hybrid Parameter-Efficient Fine-Tuning Approach for Hippocampus Segmentation and Alzheimer's Disease Diagnosis

Wangang Cheng[1], Guanghua He[1], Keli Hu[2], Mingyu Fang[3], Liang Dong[3], Zhong Li[4], Hancan Zhu[1,2*]

[1] School of Mathematics, Physics and Information, Shaoxing University, Shaoxing, Zhejiang, 312000, China

[2] Institute of Artificial Intelligence, Shaoxing University, Shaoxing, Zhejiang, 312000, China

[3] Medical School, Shaoxing University, Shaoxing, Zhejiang, 312000, China

[4] School of Information Engineering, Huzhou University, Huzhou, Zhejiang 313000, China

*Corresponding author:

Hancan Zhu, 900 ChengNan Rd, School of Mathematics Physics and Information, Shaoxing University, Shaoxing, Zhejiang, China 312000. Email: hancanzhu@yeah.net

## Abstract

Deep learning methods have significantly advanced medical image segmentation, yet their success hinges on large volumes of manually annotated data, which require specialized expertise for accurate labeling. Additionally, these methods often demand substantial computational resources, particularly for three-dimensional medical imaging tasks. Consequently, applying deep learning techniques for medical image segmentation with limited annotated data and computational resources remains a critical challenge. In this paper, we propose a novel parameter-efficient fine-tuning strategy, termed HyPS, which employs a hybrid parallel and serial architecture. HyPS updates a minimal subset of model parameters, thereby retaining the pre-trained model's original knowledge structure while enhancing its ability to learn specific features relevant to downstream tasks. We apply this strategy to the state-of-the-art SwinUNETR model for medical image segmentation. Initially, the model is pre-trained on the BraTs2021 dataset, after which the HyPS method is employed to transfer it to three distinct hippocampus datasets. Extensive experiments demonstrate that HyPS outperforms baseline methods, especially in scenarios with limited training samples. Furthermore, based on the segmentation results, we calculated the hippocampal volumes of subjects from the ADNI dataset and combined these with metadata to classify disease types. In distinguishing Alzheimer's disease (AD) from cognitively normal (CN) individuals, as well as early mild cognitive impairment (EMCI) from late mild cognitive impairment (LMCI), HyPS achieved classification accuracies of 83.78% and 64.29%, respectively. These findings indicate that the HyPS method not only facilitates effective hippocampal segmentation using pre-trained models but also holds potential for aiding Alzheimer's disease detection. Our code is publicly available[1].

## 1. Introduction

In recent years, deep learning methods have significantly advanced medical image segmentation. However, their success heavily relies on the availability of large volumes of annotated data, which poses significant challenges in the medical field. Obtaining high-quality, large-scale annotated medical images is both expensive and time-consuming. Additionally, the substantial variability in medical images, due to differences in acquisition conditions or equipment parameters, often necessitates retraining segmentation models from scratch when applied to new datasets. This considerably increases both deployment and training costs. Therefore, improving the performance of medical image segmentation models with limited training data, particularly in terms of effective model transfer, has become a pressing research challenge.

---

[1] https://github.com/WangangCheng/HyPS

Parameter-efficient fine-tuning (PEFT) methods facilitate model transfer learning by updating only a small subset of parameters. Initially developed for natural language processing, PEFT methods are broadly classified into three categories: addition methods, selective methods, and reparameterization-based methods [1]. A representative example of addition methods is the Adapter [2-4], which enhances a pre-trained model by introducing additional parameters, fine-tuning only the newly added components. Due to its sequential nature, this approach is referred to as a sequential method. Selective methods target specific layers or internal structures within the model that significantly impact performance [5, 6]. In contrast, reparameterization methods introduce additional low-rank weight matrices without directly altering the pre-trained model's weights, employing low-rank approximation to reduce the number of trainable parameters; this is known as a parallel method. Examples of reparameterization methods include LoRA [7], PiSSA [8], Fact [9], and KAdaptation [10]. The potential of combining sequential and parallel methods remains largely unexplored, and most fine-tuning techniques to date have been developed primarily for natural images [11]. In medical imaging, fine-tuning methods are predominantly based on the Segment Anything Model (SAM) [12], which focuses on transferring segmentation capabilities from natural images to medical images. However, fine-tuning techniques specifically designed for state-of-the-art (SOTA) models in medical image segmentation are still underexplored.

Building on the previous analysis, we propose a new paradigm for efficient transfer learning across diverse medical image datasets—the Hybrid Parallel and Sequential method (HyPS). The HyPS method is implemented using the SwinUNETR model [13], and it optimizes the model by adding and updating only a minimal subset of parameters, thereby significantly reducing the deployment and training overhead. Specifically, we freeze certain components of the image encoder and employ a hybrid approach that integrates PiSSA [8] and Adapter [4] techniques to update the linear layer parameters within the encoder, while keeping the decoder fully trainable. Our HyPS fine-tuning strategy updates a total of 20.0724 million parameters, which represents just 31.8% of the original model's parameters. We pre-trained the SwinUNETR model on a brain tumor dataset and conducted fine-tuning experiments on three hippocampus datasets. The HyPS method demonstrates exceptional performance in scenarios with limited annotated medical data. Experimental results indicate that with only 10 training samples, our method achieves an average improvement of 1.16% in the Dice coefficient and an average reduction of 0.532 in the HD95 metric compared to the baseline method (full tuning).

To further validate the effectiveness of the HyPS method, we extended its application to the UNETR model, and the experimental results confirmed that HyPS outperforms other parameter-efficient fine-tuning techniques. Additionally, we explored potential applications of the HyPS method by analyzing hippocampal volumes derived from fine-tuned inference results. Our findings demonstrate that HyPS is highly effective in diagnosing Alzheimer's disease (AD), early mild cognitive impairment (EMCI), late mild cognitive impairment (LMCI), and cognitively normal (CN) control groups. An earlier version of this work, the CPS method, has been accepted for presentation at the international conference PRCV2024. The differences between the HyPS method presented in this paper and the CPS method are as follows:

- The CPS method integrates the Adapter and LoRA branches in a parallel fashion, while the HyPS method combines the Adapter branch and the PiSSA branch in a sequential manner.

- In the HyPS method, the PiSSA branch uses the primary singular values and singular vectors of the pre-trained model weights for initialization, rather than relying on random initialization. This enables HyPS to directly fine-tune the most influential parts of the model, leading to enhanced performance during the fine-tuning process.
- We applied HyPS to hippocampus segmentation by fine-tuning the pre-trained SwinUNETR model on a brain tumor dataset. The experimental results show that HyPS not only outperforms the PiSSA and Adapter methods individually but also exceeds the CPS method.
- We employed the HyPS method to calculate the left and right hippocampal volumes of patients in the ADNI dataset and trained an SVM classifier using both these volumes and the patients' metadata to classify disease types. The results indicate that this classifier performs exceptionally well in diagnosing AD.

## 2. Related works

### 2.1. Parameter-Efficient Fine-Tuning for Medical Imaging

PEFT has been widely applied in natural language processing and computer vision tasks [1, 3, 4, 7, 9-11, 14-16]. The core idea of PEFT is to load a pre-trained model and update only a small portion of its parameters during training. This approach preserves the original knowledge of the trained model while optimizing it for specific tasks, achieving a balance between model performance and computational efficiency. With the growing popularity of large models, PEFT methods have also been increasingly applied to medical image segmentation tasks. For example, SAMed employs the LoRA method to fine-tune the SAM large model, extending its segmentation capability from natural images to medical images [17]. MA-SAM enhances the segmentation capability of SAM in 3D medical imaging tasks by integrating 3D Adapters into the Transformer Block of the SAM image encoder, allowing the pre-trained 2D backbone to extract three-dimensional information from the input data while leveraging the pre-trained model in the original 2D backbone [18]. Additionally, several fine-tuning strategies are based on prompt designs for SAM. For instance, SAM-Med2D [19] employs a more comprehensive set of prompts, including points, bounding boxes, and masks, specifically tailoring SAM for 2D medical images. MSA [20] integrates domain-specific medical knowledge into the SAM model using point prompts and Adapter techniques, further enhancing SAM's segmentation capabilities in the medical imaging field. While SAM is a large model based on natural images, our method in this paper focuses primarily on achieving transfer learning between medical images.

### 2.2. Deep Learning in Medical Image Segmentation

Deep learning-based methods for medical image segmentation can be broadly classified into three categories: (1) Convolutional Neural Network (CNN) methods [21-23], (2) Transformer-based methods [24-28], and (3) hybrid CNN-Transformer methods [29, 30]. CNN-based methods were the first deep learning models to be widely used for medical image segmentation. CNNs excel in capturing local image features and are particularly effective at recognizing textures and edges, offering high computational efficiency. However, due to inherent inductive biases such as locality and translational invariance, CNNs struggle to establish long-range dependencies. In contrast, Transformer-based methods capture global dependencies through self-attention mechanisms, making them especially adept at understanding the overall content and structure of images. Transformer models, when trained on large datasets, demonstrate strong generalization capabilities. However, their increased complexity often requires more computational resources and annotated data, presenting challenges for medical image segmentation tasks.

Hybrid CNN-Transformer methods aim to combine the strengths of both CNNs and Transformers. For instance, CoTr [30] introduces an efficient approach that bridges CNNs and Transformers, leveraging the feature extraction efficiency of CNNs and the global context modeling of Transformers. This collaborative framework has shown remarkable performance in 3D medical image segmentation. Similarly, TransFuse [29] proposes an architecture that fuses Transformers and CNNs in parallel, utilizing their respective strengths at different stages to enhance the accuracy and efficiency of medical image segmentation. SwinUNETR [13] is a self-supervised learning architecture for medical image segmentation, featuring a Swin-Transformer-based encoder and a CNN-based decoder. The encoder is capable of extracting multi-resolution feature representations and has demonstrated excellent performance across multiple public datasets. While hybrid methods are designed to alleviate the limitations of individual approaches, they often introduce increased computational complexity, especially when managing multiple CNN and Transformer components simultaneously. Consequently, there is an urgent need to develop model training methods that can effectively maximize the utility of limited medical imaging data, ensuring robust performance while maintaining computational efficiency.

## 2.3. Deep Learning in Hippocampal Image Segmentation and Classification

Research has demonstrated a close link between AD and the structure of the hippocampus, making precise hippocampal segmentation crucial for studying this disease [31-33]. However, the hippocampus has an irregular structure with complex boundaries and closely resembles surrounding brain tissues, making precise segmentation challenging. Recent advances in deep learning have significantly improved hippocampal segmentation. Fully Convolutional Networks (FCN) [34] enhance network flexibility and efficiency by replacing fully connected layers with convolutional layers, allowing the network to process images of various sizes. DU-Net [35] builds on FCN by incorporating a dilated dense network, generating multi-scale features that improve segmentation of high-resolution hippocampal images. HGM-cNet [36] introduces a robust cascaded deep learning framework that integrates hippocampal gray matter probability maps, providing strong stability and generalization in hippocampal segmentation. Vit U-Net [37] applies the Transformer-based ViT architecture to hippocampal segmentation, demonstrating the continuous learning capabilities of Transformers in medical imaging tasks. Deephippo [38] utilizes Squeeze-and-Excitation layers as an attention mechanism to dynamically adjust the importance of different channels in CNNs, achieving precise segmentation of hippocampal images. These advancements showcase the diverse range of deep learning approaches that can enhance the accuracy of hippocampal segmentation, which is crucial for advancing research in AD and other neurological conditions.

In hippocampal classification tasks, traditional methods have predominantly relied on template matching and manual annotation. However, recent advances in machine learning and deep learning have shown promising results. The RLBP method [39] performs hippocampal segmentation within the multi-atlas segmentation framework and analyzes hippocampal volumes to develop a classification model for diagnosing AD. Liu et al. [40] proposed a multitask model that jointly learns hippocampal segmentation and disease classification. DenseCNN [41] enhances classification performance by combining hippocampal segmentation with disease classification. It extracts deep visual features through multiple convolutional and rectification layers and integrates them with global shape features for joint training. Balasundaram et al. [42] employed a combination of CNN and ResNet50 models to classify the severity of AD, using machine learning and ensemble learning algorithms for AD detection. Furthermore, 3DRA-Net [43] developed a 3D Residual Attention Network to classify CN, MCI, and AD, demonstrating that a comprehensive approach incorporating hippocampal volume, segmentation probability maps, and radiomic features can significantly enhance the diagnosis of AD.

While the aforementioned methods perform well in hippocampal segmentation and classification tasks, they often require retraining for image data from different sources, necessitating the development of separate models from scratch for each dataset. This approach is not only computationally intensive and costly to deploy but also demands extensive manual annotations for each dataset, significantly increasing the labor burden. These challenges highlight the need for more efficient models capable of effectively transferring across diverse datasets and tasks, thereby reducing computational costs and the reliance on extensive manual annotation, making these methods more practical for real-world applications.

## 3. Methods

### 3.1. HyPS Fine-Tuning Method

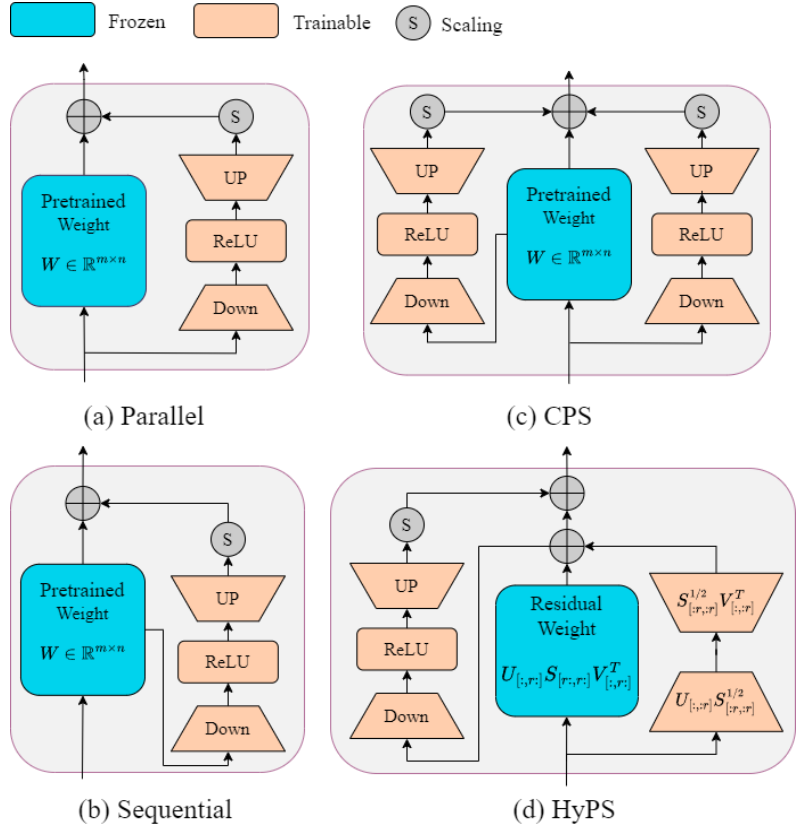

Figure 1. (a) The improved parallel method of LoRA, (b) the sequential method of SeqLoRA, (c) the CPS method, which combines parallel LoRA and sequential SeqLoRA, and (d) the HyPS method, which integrates PiSSA and SeqLoRA.

Our fine-tuning method, HyPS, is inspired by several approaches and effectively combines parallel and sequential fine-tuning techniques, including LoRA [7], Adaptformer [3], PiSSA [8], Hydra [11]. LoRA, a parallel fine-tuning method, utilizes linear adapters in the pretrained model's linear layers to enable effective model adaptation. We introduced a modification to LoRA by incorporating nonlinear activation functions and scaling factors, as shown in Figure 1(a). The modified calculation process is as follows: Given a weight matrix $A \in \mathbb{R}^{m \times n}$, it is decomposed into $A = A_{up}A_{down}$, where $A_{up} \in \mathbb{R}^{m \times r}$ and $A_{down} \in \mathbb{R}^{r \times n}$ are the projection matrices for the up and down transformations, respectively, with $r \ll \min(m, n)$. For an input feature $x \in \mathbb{R}^n$, the modified LoRA module can be expressed as:

$$x' = Wx + b + \Delta Wx = Wx + b + sA_{up}ReLU(A_{down}x), \tag{1}$$

where $x' \in \mathbb{R}^m$ represents the output vector, $W \in \mathbb{R}^{m \times n}$ is the pretrained weight, and $b \in \mathbb{R}^m$ denotes the bias. The insertion of the ReLU activation function between the up and down projections not only enhances the model's ability to represent complex features but also improves its adaptation to task-specific fine-tuning [44]. Here, $s$ is a scaling factor used to balance task-agnostic and task-specific features, with a default value of $s = 1$. To effectively optimize linear adaptability during the fine-tuning, the pretrained weight $W$ remains frozen; only the increment $\Delta W$ is updated. This parallel fine-tuning method optimizes the pretrained weights without directly modifying them, which is advantageous for learning new features that differ from those in the pretrained model. However, the low-rank matrices introduced by LoRA into the original model structure may not fully capture the distributional differences between the new training data and the pretrained model, potentially resulting in suboptimal adaptation.

SeqLoRA, the sequential counterpart to LoRA, is conceptually similar to the recently proposed RepAdapter [45]. Both SeqLoRA and RepAdapter focus on sequential linear adapter modules and serve as extensions of the Adapter framework. The structure of SeqLoRA is illustrated in Figure 1(b), and its computation is defined as follows:

$$x' = Wx + b + sB_{up}\text{ReLU}(B_{down}(Wx + b)), \qquad (2)$$

where $B_{up} \in \mathbb{R}^{m \times r}$ is the up-projection matrix, and $B_{down} \in \mathbb{R}^{r \times m}$ is the down-projection matrix. SeqLoRA learns new features specific to downstream tasks by leveraging the linear combination of pretrained features, effectively reducing the risk of overfitting to the pretrained model's generalization ability. However, SeqLoRA might face challenges when learning new task-specific features that differ significantly from those in the pretrained stage, as it may not fully capture the distinct characteristics of the pretrained data, potentially leading to suboptimal transfer across domains with diverse data distributions.

To fully exploit the advantages of both LoRA and SeqLoRA, we previously proposed the CPS method, as shown in Figure 1(c). CPS combines the parallel structure of LoRA with the sequential structure of SeqLoRA, providing a more flexible fine-tuning strategy. The computation process is defined as follows:

$$x' = Wx + b + s_a A_{up}\text{ReLU}(A_{down}x) + s_b B_{up}\text{ReLU}(B_{down}(Wx + b)), \qquad (3)$$

where $A_{up} \in \mathbb{R}^{m \times r_a}$ and $A_{down} \in \mathbb{R}^{r_a \times n}$ represent the low-rank adaptation matrices for the parallel branch with rank $r_a$, and $s_a$ is the scaling factor for this branch. Similarly, $B_{up} \in \mathbb{R}^{m \times r_b}$ and $B_{down} \in \mathbb{R}^{r_b \times m}$ represent the low-rank adaptation matrices for the sequential branch with rank $r_b$, and $s_b$ is the corresponding scaling factor. For simplicity, we set $r_a = r_b$ and $s_a = s_b = 1$. The up-projection matrices $A_{up}, B_{up}$ are initialized to zero, while the down-projection matrices $A_{down}, B_{down}$ are set using Kaiming initialization [46]. All these matrices $A_{up}, A_{down}, B_{up}, B_{down}$ are updated through stochastic gradient descent, with the pretrained weights $W$ and bias $b$ kept frozen. This dual approach enables the pretrained model to adapt more effectively to new tasks while preserving its original knowledge structure. Consequently, the model can efficiently learn new features without compromising its generalization capabilities.

In this paper, we propose the HyPS fine-tuning method, an enhancement of the CPS method, as illustrated in Figure 1(d). Inspired by the PiSSA approach [8], we first apply Singular Value Decomposition (SVD) to the pretrained weights $W \in \mathbb{R}^{m \times n}$, resulting in:

$$W = USV^T,$$

where $U \in \mathbb{R}^{m \times min\{m,n\}}$, $S \in \mathbb{R}^{min\{m,n\} \times min\{m,n\}}$, and $V \in \mathbb{R}^{n \times min\{m,n\}}$. We then decompose $W$ into a primary matrix $W^{Pri}$ and a residual matrix $W^{Res}$ such that:

$$W = W^{Res} + W^{Pri}.$$

The primary matrix is computed as follows:

$$W^{Pri} = W_{up}^{Pri} W_{down}^{Pri},$$

where

$$W_{up}^{Pri} = U_{[:,:r]} S_{[:r,:r]}^{1/2} \in \mathbb{R}^{m \times r}, \text{ and } W_{down}^{Pri} = S_{[:r,:r]}^{1/2} V_{[:,:r]}^{T} \in \mathbb{R}^{r \times n}.$$

In the above equations, $r$ is a hyperparameter, typically satisfying $r \ll \min(m,n)$. The residual matrix is given by:

$$W^{Res} = U_{[:,r:]} S_{[r:,r:]} V_{[:,r:]}^{T}.$$

After obtaining these matrices, we combine them with the sequential adapter SeqLoRA. The HyPS computation is as follows:

$$x' = sB_{up} \text{ReLU}\left(B_{down}\left((W^{Res} + W^{Pri})x + b\right)\right) + (W^{Res} + W^{Pri})x + b. \tag{4}$$

The HyPS method effectively integrates the PiSSA approach with SeqLoRA by leveraging SVD to decompose $W$ into a primary matrix $W^{Pri}$ and a residual matrix $W^{Res}$. The primary matrix $W^{Pri}$ captures the dominant singular values and their associated singular vectors, representing the most influential components of $W$ within the model, which are set to be updated during training. In contrast, the residual matrix $W^{Res}$ comprises the less significant singular values and their corresponding singular vectors, which have a reduced impact on the model and remain frozen during training. We initialize the primary components $W^{Pri}$ with a low-rank approximation of the pretrained weights, allowing the model to adapt in the most critical directions during training. In the SeqLoRA branch, the up and down projection matrices are initialized with zeros and Kaiming random values, respectively, and $W^{Pri}, B_{up}$, and $B_{down}$ are updated through stochastic gradient descent. The rank of these matrices is a crucial hyperparameter, and for simplicity, we set the ranks of the sequential and parallel branches to be identical.

### 3.2. Training Strategy

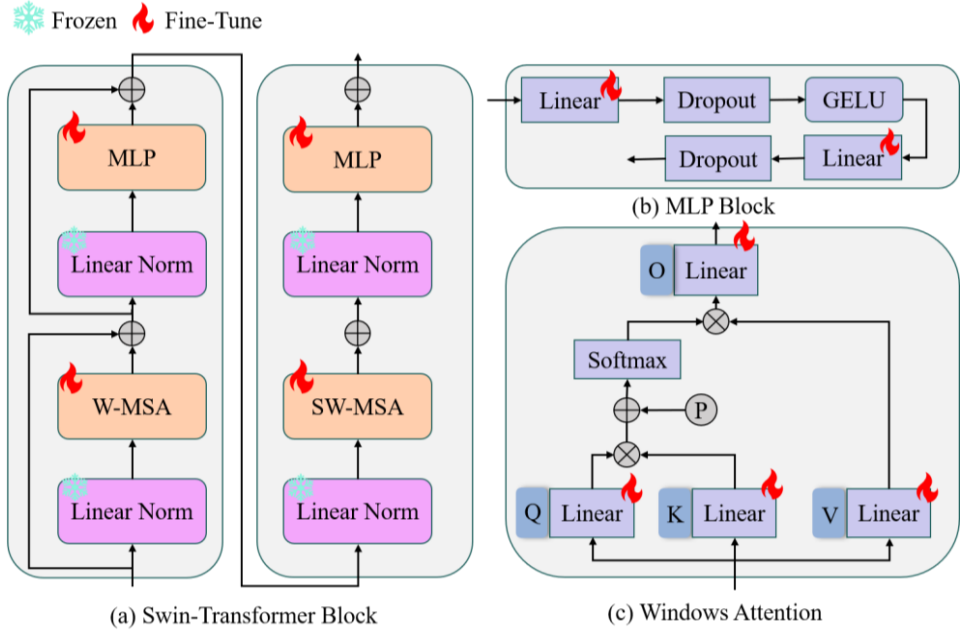

Figure 2. Key Modules in SwinUNETR: (a) Schematic of the Swin-Transformer Block, (b) MLP Block, (c) Window Attention Mechanism in W-MSA and SW-MSA.

We applied the HyPS method to the pre-trained SwinUNETR model [13] for parameter fine-tuning, facilitating transfer learning by updating only a small portion of the model's parameters. SwinUNETR features a U-shaped architecture composed of an encoder and a decoder. The encoder is built around

Swin-Transformer Blocks, which capture multi-scale contextual information, while the decoder uses convolutional neural networks to reconstruct the segmentation map. The model integrates feature maps from the encoder with corresponding layers in the decoder through skip connections, enhancing spatial information retention. This architecture has demonstrated exceptional performance in medical image segmentation tasks.

We initially pre-trained the SwinUNETR model on the BraTS 2021 dataset [47], adhering to its original network architecture. Following this, we employed our proposed HyPS method to perform parameter-efficient transfer learning on the hippocampus dataset. The pre-trained model already encapsulates substantial imaging background knowledge, which is effectively harnessed during transfer learning. Specifically, HyPS was applied to the linear layers within the Swin-Transformer Block, including the two linear layers in the MLP block and the linear layers used for computing the Q (query), K (key), V (value), and O (output) matrices in the Window Attention module, as illustrated in Figure 2. During fine-tuning, we updated only the newly introduced parameters and the weights of the convolutional layers in the decoder, while keeping all other weights frozen. The decoder was not frozen because, in medical imaging, many anatomical structures or lesions are small and require higher resolution for accurate differentiation. The decoder plays a crucial role in restoring high-resolution features, which is essential for precise segmentation in medical imaging.

## 4. Hippocampal Segmentation: Experimental Results and Analysis

### 4.1. Dataset and Preprocessing

**The BraTS 2021 dataset** [47] is a large-scale, multi-modal brain MRI dataset containing 8,160 MRI scans from 2,040 patients. Each patient's data includes four MRI modalities: T1, T1ce, T2, and FLAIR, with each modality having a resolution of 240×240×155 and a voxel size of 1×1×1 mm³. The dataset's primary annotations include enhancing tumor (ET), tumor-associated edema/invaded tissue (ED), and necrotic tumor core (NCR), with ground truth labels available for 1,251 cases. We utilized these 1,251 publicly available annotated T1ce modality scans for pre-training our model, dividing the data into training, validation, and test sets with a ratio of 1000:125:126. For training, the three annotated regions were merged into a single region. Following the parameter settings of SwinUNETR [13], we trained the model for 400 epochs. The pre-trained model achieved a Dice coefficient of 81.9% on the BraTS 2021 dataset, indicating strong performance in brain tumor segmentation.

**The EADC dataset** [48] obtained from the ADNI database[2], includes 135 MRI scans with hippocampal annotations. Each scan has a resolution of 197×233×189 voxels, with a voxel size of 1×1×1 mm³. After identifying five scans with annotation inconsistencies, these were excluded, leaving a final dataset of 130 scans.

**The LPBA40 dataset** [49] comprises 3D brain MRI scans from 40 healthy adults, with detailed annotations for 56 distinct brain tissues. In our study, we concentrated on the hippocampal region. Each scan has a resolution of 256×124×256 voxels, with a voxel size of 0.8938×1.500×0.8594 mm³.

**The HFH dataset** [50] contains 50 T1-weighted brain MRI scans with hippocampal labels, acquired using MRI machines from two different manufacturers, leading to variations in resolution and contrast. The scans have dimensions of 256×124×256 and 512×124×512 voxels, with corresponding voxel sizes of 0.781×2.000×0.781 mm³ and 0.39×2.00×0.39 mm³. Due to the availability of precise annotations for only 25 scans, our experiments were conducted using this subset.

---

[2] http://adni.loni.usc.edu/

For the three hippocampus datasets mentioned above, we used FSL[3] to align all scans to the MNI152 standard space [51]. Standard preprocessing steps were then applied, including skull stripping, resampling, and affine transformation. The resulting images were standardized to a resolution of 182×218×182 voxels with a voxel size of 1×1×1 mm³.

## 4.2. Training and Inference Details

The fine-tuning experiments for the SwinUNETR model were conducted using the Adam optimizer with a batch size of 4. All experiments were implemented in the PyTorch framework and executed on two NVIDIA GeForce RTX 4090D GPUs. We employed a polynomial learning rate schedule, starting with an initial learning rate of 0.001 and decaying it by a factor of 0.9 at each iteration. Image patches of size 128×128×128 were randomly cropped from the dataset and used as inputs for the network. To enhance data variability, several image augmentation strategies were applied, including: (1) random mirror flipping with a 50% probability along the axial, coronal, and sagittal planes; (2) random intensity shifts within the range of [-0.1, 0.1], where a random value is added to or subtracted from each pixel's intensity; and (3) random scaling, where the image is resized within a range of [0.9, 1.1]. The network's loss function is the Dice loss [22], which is defined as:

$$\mathcal{L}(Y, \tilde{Y}) = -\frac{1}{N} \sum_{n=1}^{N} \frac{2 Y_n \tilde{Y}_n}{Y_n + \tilde{Y}_n}, \quad (5)$$

where, $Y_n$ and $\tilde{Y}_n$ represent the ground truth and the predicted probability, respectively, while $N$ denotes the batch size. The loss function includes an L2 regularization term to prevent overfitting, with a weight decay rate of $10^{-5}$. The network was trained for a total of 1,000 epochs, after which training was stopped.

During the testing phase, image patches of size 128×128×128 were extracted and input into the trained model using a non-overlapping sliding window strategy for segmentation. The final inference result was obtained by averaging the model outputs from the last four epochs. In the post-processing step, false positives in the segmented hippocampus were removed. This process involved generating a binary mask to identify all connected target regions and applying a threshold of 1,000 voxels (equivalent to 1 cm³). Any connected regions smaller than this threshold were reclassified as background. This post-processing technique effectively reduces model errors and noise, enhancing overall segmentation performance.

We selected the Dice coefficient and the 95th percentile of the Hausdorff Distance (Hausdorff Distance 95%) as evaluation metrics. The Dice coefficient measures the relative overlap volume between the automatic segmentation and the ground truth, defined as:

$$Dice = 2 \frac{V(A \cap B)}{V(A) + V(B)} \times 100\%, \quad (6)$$

where $A$ represents the ground truth, and $B$ denotes the automatic segmentation. $V(S)$ indicates the volume of $S$. The Hausdorff Distance 95% is a robust version of the Hausdorff Distance, evaluating the segmentation structure's robustness and boundary consistency. It is defined as:

$$HD_{95}(A, B) = \max(h_{95}(A, B), h_{95}(B, A)), \quad (7)$$

where $h_{95}(A, B) = K_{a \in A}^{95} \min_{b \in B} d(a, b)$ is the 95th percentile of the minimum Euclidean distances between boundary points in $A$ and $B$.

---

[3] https://fsl.fmrib.ox.ac.uk/fsl/fslwiki/

## 4.3. Results and Analysis of Segmentation Performance

Table 1．Impact of Rank Selection on the Performance of Various Fine-Tuning Approaches with the EADC Dataset. Higher Dice coefficients and lower HD95 values indicate better segmentation. The best Dice values for each method are highlighted in bold.

|  | r=2 | | r=4 | | r=8 | | r=16 | | R=32 | |
| --- | --- | --- | --- | --- | --- | --- | --- | --- | --- | --- |
|  | Dice | HD95 | Dice | HD95 | Dice | HD95 | Dice | HD95 | Dice | HD95 |
| LoRA | 86.32 | 4.600 | 86.53 | 4.495 | **86.54** | 4.462 | 86.30 | 4.725 | 86.37 | 4.596 |
| SeqLoRA | 86.07 | 4.767 | 85.87 | 4.793 | 86.26 | 4.758 | **86.32** | 4.604 | 86.13 | 4.523 |
| PiSSA | 86.48 | 4.736 | 86.76 | 4.511 | 86.76 | 4.576 | 86.72 | 4.420 | **86.83** | 4.460 |
| CPS | 86.19 | 4.454 | 86.87 | 4.784 | **87.02** | 4.512 | 86.32 | 4.763 | 86.33 | 4.533 |
| HyPS | 87.29 | 4.204 | 87.47 | 4.120 | **87.60** | 4.052 | 86.56 | 4.354 | 86.73 | 4.477 |

We adapted the SwinUNETR model, originally pre-trained on the BraTS 2021 dataset, for application to hippocampus datasets. For each hippocampus dataset, 10 subjects were randomly selected for training, while the remaining subjects were reserved for testing during fine-tuning experiments. Given the significant impact of rank selection on fine-tuning performance, we first evaluated the effects of various rank values on segmentation accuracy using the EADC dataset, as shown in Table 1. Our proposed HyPS method achieved the highest Dice coefficient and the lowest HD95 at a rank of r=8. In contrast, the PiSSA and SeqLoRA methods required higher ranks to reach their peak accuracy, which also led to increased computational demands. The HyPS method demonstrated that optimal accuracy could be achieved with a lower rank, thereby improving computational efficiency. For each fine-tuning method, we determined the rank that yielded the highest Dice coefficient as the optimal rank. All subsequent experiments were conducted using these optimal values to ensure a fair comparison across different methods.

Table 2．Comparison of Segmentation Results from Fine-Tuning Strategies on SwinUNETR. Higher Dice coefficients and lower HD95 values indicate better segmentation performance, with the best results highlighted in bold.

|  |  | EADC | | LPBA40 | | HFH | | | |
| --- | --- | --- | --- | --- | --- | --- | --- | --- | --- |
| Method | Params(M) | Dice | HD95 | Dice | HD95 | Dice | HD95 | Avg_Dice | Avg_HD95 |
| Full tuning | 63.1279 | 85.93 | 4.774 | 83.85 | 6.354 | 84.89 | 5.033 | 84.89 | 5.387 |
| Linear-probing | 22.7384 | 86.03 | 4.643 | 83.56 | 6.401 | 84.25 | 5.449 | 84.61 | 5.498 |
| LoRA | 19.8398 | 86.54 | 4.462 | 83.70 | 6.271 | 84.23 | 4.731 | 84.82 | 5.155 |
| SeqLoRA | 20.0586 | 86.26 | 4.758 | 83.52 | 6.422 | 84.52 | 4.991 | 84.76 | 5.390 |
| SSF | 19.6308 | 85.50 | 4.794 | 82.77 | 6.484 | 83.09 | 5.964 | 83.79 | 5.747 |
| PiSSA | 20.3671 | 86.65 | 4.635 | 83.79 | 6.093 | 84.83 | 4.551 | 85.09 | 5.093 |
| CPS | 20.0701 | 87.02 | 4.512 | **85.04** | 6.076 | 85.51 | 4.714 | 85.86 | 5.101 |
| HyPS | 20.0724 | **87.60** | **4.052** | 84.89 | **5.869** | **85.66** | **4.645** | **86.05** | **4.855** |

We evaluated the segmentation accuracy of various fine-tuning methods across three hippocampus datasets, as summarized in Table 2. Full tuning involves updating all model parameters, whereas Linear-probing focuses on updating only the linear layers within the Swin-Transformer Block. The SSF method [1] introduces scale and rotation factors after each Transformer operation (including multi-head attention, MLP, LN, etc.) and updates only these additional parameters, keeping the rest of the model frozen. This approach is applied to the model's encoder, similar to HyPS, while the decoder remains fully trainable. In Table 2, the parameter counts for PETL methods reflect the number of trainable parameters, with the model's decoder alone accounting for 19.5978 million parameters. In contrast, the trainable parameter count in our method's encoder is only 0.4746 million, significantly enhancing computational efficiency. Despite the reduced parameter count, our method achieves high-precision hippocampus segmentation. Using full tuning as the baseline, our method improved the average Dice score by 1.16% and reduced the average HD95 by 0.532 across the three hippocampus datasets. Compared to the LoRA method, our approach enhanced the average Dice coefficient by 1.23% and reduced the average HD95 by 0.3.

Additionally, our method outperforms other state-of-the-art fine-tuning techniques. The visual segmentation results in Figure 3 further demonstrate that the HyPS fine-tuning method provides richer boundary information and minimizes segmentation redundancies compared to other methods.

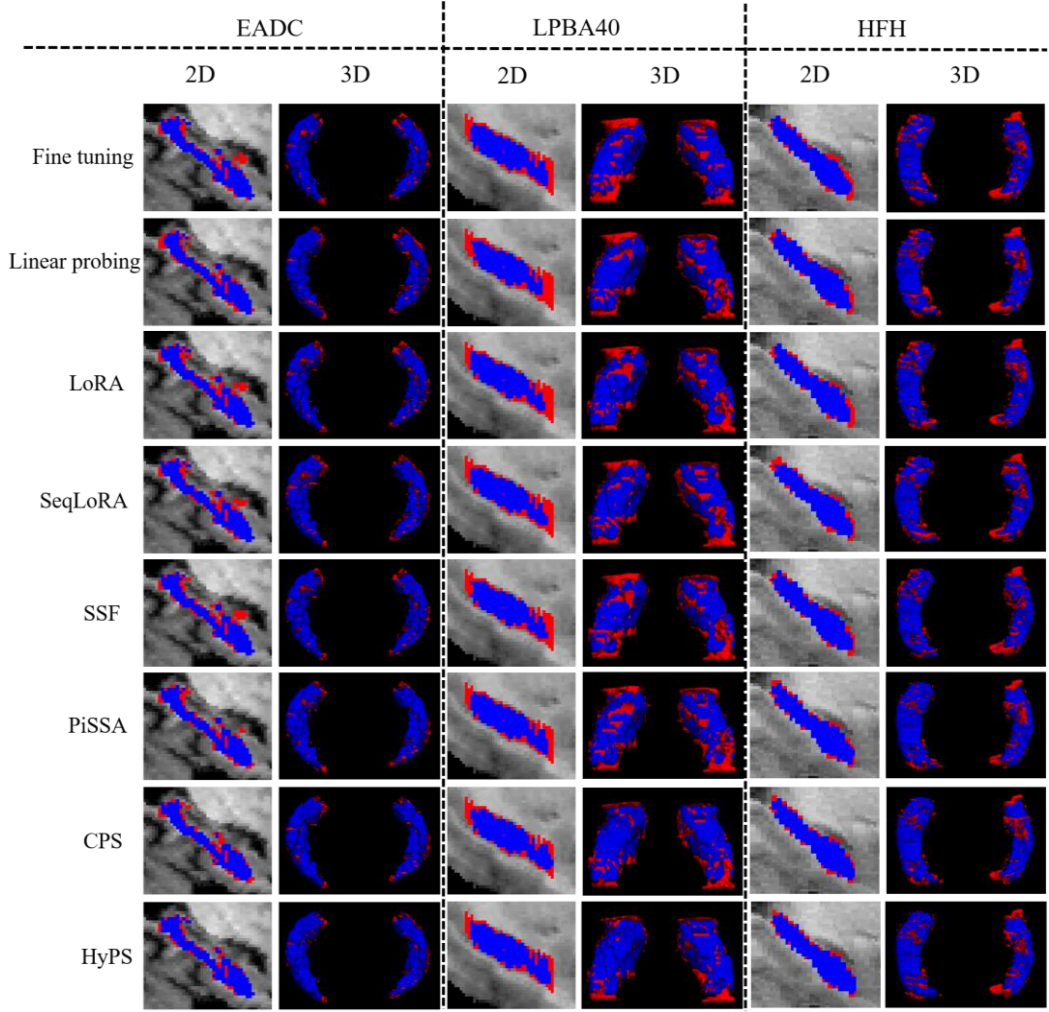

Figure 3. Visual Comparison of Segmentation Outcomes from Fine-Tuning Methods on Three Hippocampus Datasets. Each panel displays segmentation results for a randomly selected individual, including 2D slice images and 3D surface renderings. Blue regions indicate the overlap between automatic segmentation and ground truth, while red regions highlight discrepancies.

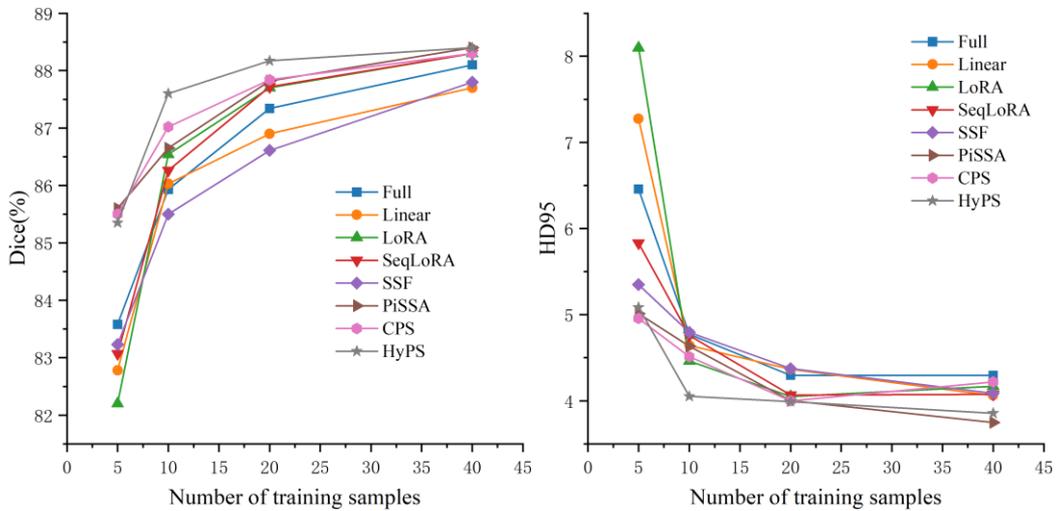

Figure 4．Comparative Analysis of Fine-Tuning Strategies on the EADC Dataset with Varying Training Data Sizes.

To investigate the impact of training data size on parameter-efficient fine-tuning methods, we conducted additional experiments on the EADC dataset, with the results presented in Figure 4. The figure shows that as the training data size increases, the performance gap between parameter-efficient methods and full fine-tuning narrows. Notably, the HyPS method consistently achieves the highest Dice score across all data sizes compared to other fine-tuning approaches. Moreover, when training data is limited, the HyPS method demonstrates a significant advantage in segmentation performance over other parameter-efficient techniques. This is particularly beneficial in medical image segmentation, where large annotated datasets are often scarce.

Our fine-tuning method is not restricted to the SwinUNETR model; it can be applied to any medical image segmentation model based on the Transformer architecture. To validate the generalizability of the HyPS method, we applied it to the UNETR model [28], which features an encoder primarily composed of a Vision Transformer (ViT) [52] structure. As with SwinUNETR, we fine-tuned the linear layers within the ViT, including the two linear layers in the MLP block and the linear layers responsible for computing Q (query), K (key), V (value), and O (output) in the self-attention block, while keeping the remaining parameters frozen and the decoder parameters trainable. We maintained the same experimental setup by randomly selecting 10 images from each hippocampus dataset for training and using the remaining images for testing.

The results of different fine-tuning methods are presented in Table 3. Notably, the UNETR model's decoder consists of only 2.6785 million parameters, and with the HyPS fine-tuning method, the number of trainable parameters is reduced by 94.42% compared to full tuning. Despite this significant reduction, the HyPS method consistently achieved the highest Dice coefficient across all three datasets, demonstrating that our fine-tuning strategy is equally effective when applied to the UNETR model. This underscores the versatility and effectiveness of our approach across different Transformer-based models in medical image segmentation.

Table 3. Comparison of Segmentation Results from Fine-Tuning Strategies on UNETR. Higher Dice coefficients and lower HD95 values indicate better segmentation, with the best results highlighted in bold.

| Method | Params(M) | EADC Dice | EADC HD95 | LPBA40 Dice | LPBA40 HD95 | HFH Dice | HFH HD95 | Avg_Dice | Avg_HD95 |
|---|---|---|---|---|---|---|---|---|---|
| Full tuning | 93.0112 | 84.74 | 5.183 | 82.37 | 6.481 | 83.24 | **4.966** | 83.45 | 5.543 |
| Linear-probing | 59.3477 | 84.85 | 4.916 | 81.36 | 6.640 | 83.17 | 5.303 | 83.13 | 5.620 |
| LoRA | 3.8581 | 85.08 | 5.242 | 82.45 | 6.546 | 81.83 | 5.769 | 83.12 | 5.852 |
| SeqLoRA | 5.3327 | 84.76 | 5.110 | 81.88 | 6.461 | 82.69 | 5.559 | 83.11 | 5.710 |
| SSF | 2.8828 | 84.79 | 5.034 | 80.69 | 7.090 | 81.57 | 5.271 | 82.35 | 5.798 |
| PiSSA | 7.4462 | 86.12 | 4.625 | 82.39 | 6.724 | 83.37 | 5.649 | 83.96 | 5.666 |
| CPS | 5.1852 | 85.54 | 5.031 | 83.08 | 6.663 | 82.59 | 5.714 | 83.74 | 5.802 |
| HyPS | 5.1883 | **86.41** | **4.497** | **83.89** | **6.410** | **84.33** | 4.995 | **84.88** | **5.301** |

## 5. Hippocampal Volume Analysis and AD Diagnosis

By leveraging PEFT techniques, clinicians can rapidly deploy hippocampus segmentation models with minimal data annotation, facilitating their application in AD diagnosis. In this section, we conducted two classification experiments: AD vs. CN and EMCI vs. LMCI. We used the left and right hippocampal volumes derived from segmentation results produced by various fine-tuning methods, along with patient meta-information such as age and gender, to train a Support Vector Machine (SVM) classifier. This approach not only evaluates the effectiveness and accuracy of the segmentation techniques but also aims to enhance the precision of early AD diagnosis, providing a more reliable auxiliary tool for clinical practice.

## 5.1. Data Description

We conducted experiments on the ADNI dataset[4], randomly selecting T1-weighted MRI data from 261 subjects in the ADNI-2 cohort. To create hippocampus segmentation labels for training, we first used the wholeBrainSeg-Large UNEST segmentation model from the MONAI Bundle app in 3D Slicer to segment 10 randomly selected images. These segmentations were then meticulously refined under the guidance of two expert radiologists, resulting in high-quality final segmentation labels. These 10 images constituted the training set for our segmentation model, while the remaining 251 images were reserved for the classification task, which included four categories: CN, AD, EMCI, and LMCI (details in Table 4). All images underwent preprocessing with FreeSurfer[5] including bias field correction and brain extraction [53]. Additionally, FSL software was used to register all images to the MNI152 standard space, resulting in the images being resized to dimensions of 182×218×182 with a voxel size of 1×1×1 mm³.

Table 4. Demographic and Diagnostic Details of 251 Selected Subjects from the ADNI-2 Cohort Used for AD Diagnosis.

|  | CN | AD | EMCI | LMCI |
| --- | --- | --- | --- | --- |
| Number of subjects | 66 | 46 | 78 | 61 |
| Age (years): mean±std | 75.20±7.07 | 73.91±8.31 | 71.99±7.19 | 73.38±8.40 |
| Males/Females | 37/29 | 24/22 | 45/33 | 29/32 |

## 5.2. Hippocampal Volume Measurement

To calculate hippocampal volumes segmented by different fine-tuning methods, we used 10 ADNI images with expert annotations as the training set for parameter fine-tuning, ensuring consistency with the parameter settings from previous SwinUNETR experiments. The trained model was then applied to segment the remaining 251 unannotated images. To remove isolated points in the segmentation results, we employed the same post-processing technique described earlier. Subsequently, we calculated the left and right hippocampal volumes for all subjects, with the results summarized in Table 5. The data reveals a clear trend: $V_{AD} < V_{EMCI} < V_{LMCI} < V_{CN}$, where $V$ represents the hippocampal volume. This progression indicates that hippocampal volume decreases from CN to AD, with volumes for EMCI and LMCI occupying intermediate positions, aligning with clinical expectations.

Table 5. Left and Right Hippocampal Volumes of Four Different Groups of Subjects Segmented by Different Fine-Tuning Methods (mean ± std, cm³).

|  |  | AD | CN | EMCI | LMCI |
| --- | --- | --- | --- | --- | --- |
| Full tuning | Left | 2.282±0.138 | 2.699±0.104 | 2.593±0.149 | 2.408±0.194 |
|  | Right | 2.199±0.132 | 2.763±0.145 | 2.663±0.190 | 2.421±0.212 |
| Linear-probing | Left | 2.296±0.116 | 2.724±0.121 | 2.634±0.122 | 2.412±0.194 |
|  | Right | 2.146±0.149 | 2.745±0.151 | 2.643±0.204 | 2.392±0.229 |
| LoRA | Left | 2.298±0.143 | 2.707±0.095 | 2.607±0.148 | 2.406±0.175 |
|  | Right | 2.217±0.107 | 2.690±0.123 | 2.613±0.169 | 2.396±0.179 |
| SeqLoRA | Left | 2.281±0.156 | 2.738±0.099 | 2.639±0.156 | 2.407±0.199 |
|  | Right | 2.217±0.147 | 2.757±0.133 | 2.701±0.188 | 2.446±0.221 |
| SSF | Left | 2.351±0.137 | 2.787±0.107 | 2.698±0.148 | 2.469±0.208 |
|  | Right | 2.234±0.102 | 2.769±0.131 | 2.675±0.178 | 2.450±0.192 |
| PiSSA | Left | 2.177±0.253 | 2.713±0.124 | 2.600±0.176 | 2.337±0.285 |
|  | Right | 2.115±0.172 | 2.721±0.162 | 2.639±0.213 | 2.375±0.262 |
| CPS | Left | 2.269±0.137 | 2.706±0.110 | 2.611±0.158 | 2.386±0.205 |
|  | Right | 2.237±0.112 | 2.730±0.124 | 2.662±0.185 | 2.430±0.201 |
| HyPS | Left | 2.372±0.155 | 2.822±0.108 | 2.732±0.178 | 2.501±0.220 |
|  | Right | 2.232±0.108 | 2.755±0.163 | 2.689±0.197 | 2.457±0.231 |

---

[4] http://adni.loni.usc.edu/
[5] https://surfer.nmr.mgh.harvard.edu/

## 5.3. Classification and Estimation Performance Metrics

We developed a Support Vector Machine (SVM) classifier using the left and right hippocampal volumes, along with the subjects' gender and age information. The classifier was configured with the following default parameters: C = 1, gamma = 'auto', and kernel = 'rbf'. To ensure robustness and generalizability, we employed a five-fold cross-validation strategy during model training. The classification performance was assessed using several key metrics: Precision, Sensitivity, Specificity, F1-score, Accuracy, and Area Under the ROC Curve (AUC).

- **Precision**: The proportion of true positive predictions among all positive predictions made by the model, indicating how accurately the model identifies actual positive cases. It is defined as:

$$Precision\ =\ \frac{TP}{TP + FP}$$

where $TP$ is the number of true positives, and $FP$ is the number of false positives.

- **Sensitivity**: The proportion of actual positive cases correctly identified by the model. It measures the model's ability to detect positive cases. It is defined as:

$$Sensitivity\ =\ \frac{TP}{TP + FN}$$

where $FN$ is the number of false negatives.

- **Specificity:** The proportion of actual negative cases correctly identified by the model. It measures the model's ability to detect negative cases. It is defined as:

$$Specificity\ =\ \frac{TN}{TN + FP}$$

where $TN$ is the number of true negatives.

- **F1-score**: The harmonic mean of precision and sensitivity, balancing the contribution of both metrics. It is particularly useful when the class distribution is imbalanced. It is defined as:

$$F1 - score\ =\ \frac{2 * Precision * Sensitivity}{Precision + Sensitivity}.$$

- **Accuracy**: The proportion of all correct predictions (both true positives and true negatives) out of the total number of predictions made by the model. It is defined as:

$$Overall\ Accuracy\ =\ \frac{TP + TN}{TP + TN + FP + FN}.$$

- **Area Under the ROC Curve (AUC)**: A metric that quantifies the overall performance of a binary classifier by measuring the area under the Receiver Operating Characteristic (ROC) curve. The ROC curve plots the true positive rate against the false positive rate at various threshold settings. A higher AUC indicates better performance of the model in distinguishing between positive and negative classes.

## 5.4. Classification Analysis and Results

The results presented in Table 6 and Table 7 assess the effectiveness of segmentation outcomes, combined with additional subject characteristics, in distinguishing between diagnostic categories such as AD vs. CN and EMCI vs. LMCI. The findings indicate that the HyPS method achieved the highest Accuracy in both cases, with 83.78% for AD vs. CN and 64.29% for EMCI vs. LMCI, reflecting improvements of 2.7% and 0.72%, respectively, over the full tuning method. Additionally, the HyPS method surpassed other PEFT methods across various classification metrics. The significant hippocampal volume difference between AD patients, who exhibit marked atrophy, and CN subjects facilitates easier classification by the SVM. In contrast, the smaller hippocampal volume difference between EMCI and LMCI subjects poses a greater challenge for classification. Nonetheless, the HyPS method consistently demonstrated high accuracy compared to other PEFT methods, highlighting its robustness and effectiveness in clinical diagnostic tasks.

Table 6. Evaluation of SVM Performance for AD vs. CN Classification Using Left and Right Hippocampal Volumes from the SwinUNETR Model with Various Fine-Tuning Methods, Combined with Age and Gender. Higher values for metrics such as AUC, Precision, Sensitivity, Specificity, F1-score, and Accuracy indicate better performance. The best results are highlighted in bold.

| Method | AUC | Precision | Sensitivity | Specificity | F1-score | Accuracy |
| --- | --- | --- | --- | --- | --- | --- |
| Full tuning | 0.8929 | 80.36% | 80.91% | 80.00% | 80.57% | 81.08% |
| Linear-probing | 0.8943 | 80.44% | 81.26% | 82.22% | 80.68% | 81.08% |
| LoRA | 0.8906 | 81.67% | 82.73% | 86.67% | 81.73% | 81.98% |
| SeqLoRA | 0.8919 | 81.42% | 82.37% | 84.44% | 81.65% | 81.98% |
| SSF | 0.8818 | 79.81% | 80.86% | 84.44% | 79.90% | 80.18% |
| PiSSA | **0.8973** | 82.26% | 83.13% | 84.44% | 82.52% | 82.88% |
| CPS | 0.8949 | 83.24% | 84.24% | 86.67% | 83.48% | **83.78%** |
| HyPS | 0.8946 | **83.43%** | **84.60%** | **88.89%** | **83.56%** | **83.78%** |

Table 7. Evaluation of SVM Performance for EMCI vs. LMCI Classification Using Left and Right Hippocampal Volumes from the SwinUNETR Model with Various Fine-Tuning Methods, Combined with Age and Gender. Higher values for metrics such as AUC, Precision, Sensitivity, Specificity, F1-score, and Accuracy indicate better performance. The best results are highlighted in bold.

| Method | AUC | Precision | Sensitivity | Specificity | F1-score | Accuracy |
| --- | --- | --- | --- | --- | --- | --- |
| Full tuning | 0.6793 | 62.95% | 62.34% | **73.08%** | 62.37% | 63.57% |
| Linear-probing | 0.6780 | 60.25% | 60.28% | 64.1% | 60.26% | 60.71% |
| LoRA | 0.6782 | 61.44% | 61.06% | 70.51% | 61.09% | 62.14% |
| SeqLoRA | 0.6776 | 62.19% | 61.54% | **73.08%** | 61.53% | 62.86% |
| SSF | 0.6743 | 59.58% | 59.64% | 62.82% | 59.60% | 60.00% |
| PiSSA | 0.6717 | 62.29% | 62.20% | 67.95% | 62.23% | 62.86% |
| CPS | 0.6855 | 63.81% | 63.81% | 67.95% | 63.81% | **64.29%** |
| HyPS | **0.6884** | **63.90%** | **63.98%** | 66.67% | **63.92%** | **64.29%** |

## 6. Conclusion

In this paper, we propose HyPS, a parameter-efficient fine-tuning method specifically designed to enhance the transfer learning capabilities of medical imaging segmentation models. By integrating PiSSA and SeqLoRA, HyPS creates a streamlined structure that excels in generalization, computational efficiency, and versatility, making it applicable to any linear layer. We demonstrated the effectiveness of HyPS by fine-tuning a SwinUNETR model pre-trained on a brain tumor dataset for hippocampus segmentation. Our results show that HyPS significantly enhances performance in scenarios with limited training samples and is adaptable across other Transformer-based medical image segmentation frameworks.

HyPS addresses a critical challenge in medical image segmentation: the difficulty of acquiring large-scale, high-quality annotated data. Its successful implementation improves the accuracy and efficiency of medical diagnostics. Using the fine-tuned segmentation model, we calculated hippocampal volumes and trained SVM classifiers based on these volumes and patient meta-information to distinguish between AD and CN, as well as EMCI and LMCI. Our findings demonstrate that HyPS not only reduces the reliance on extensive annotated data but also provides more reliable information for subsequent diagnostic tasks, making it a promising tool for clinical applications.

## Competing Interests

The authors declare that they have no competing interests.

## Acknowledgments

This work was supported by Humanities and Social Science Fund of Ministry of Education of China (23YJAZH232), Scientific Research Project of Shaoxing University (20210038).

Data collection and sharing for this project were funded by the Alzheimer's Disease Neuroimaging Initiative (ADNI) (National Institutes of Health Grant U01 AG024904) and the Department of Defense Alzheimer's Disease Neuroimaging Initiative (DOD ADNI) (Department of Defense award number W81XWH-12-2-0012). ADNI is supported by the National Institute on Aging (NIA), the National Institute of Biomedical Imaging and Bioengineering (NIBIB), and through generous contributions from the following organizations: AbbVie, Alzheimer's Association, Alzheimer's Drug Discovery Foundation, Araclon Biotech, BioClinica, Inc., Biogen, Bristol-Myers Squibb Company, CereSpir, Inc., Cogstate, Eisai Inc., Elan Pharmaceuticals, Inc., Eli Lilly and Company, EuroImmun, F. Hoffmann-La Roche Ltd and its affiliated company Genentech, Inc., Fujirebio, GE Healthcare, IXICO Ltd., Janssen Alzheimer Immunotherapy Research & Development, LLC., Johnson & Johnson Pharmaceutical Research & Development LLC., Lumosity, Lundbeck, Merck & Co., Inc., Meso Scale Diagnostics, LLC., NeuroRx Research, Neurotrack Technologies, Novartis Pharmaceuticals Corporation, Pfizer Inc., Piramal Imaging, Servier, Takeda Pharmaceutical Company, and Transition Therapeutics. The Canadian Institutes of Health Research (CIHR) provides funding to support ADNI clinical sites in Canada. Contributions from the private sector are facilitated by the Foundation for the National Institutes of Health (FNIH). The Northern California Institute for Research and Education (NCIRE) serves as the grantee organization, and the study is coordinated by the Alzheimer's Therapeutic Research Institute (ATRI) at the University of Southern California (USC). Data from the ADNI study are disseminated by the Laboratory for Neuro Imaging (LONI) at the University of Southern California.